# Comparative Performance Analysis of NIST PQC Standards: From STM32 Software Limitations to FPGA-SoC Acceleration

Mustafa Akif Yildirim[1*], Osman Tokluoglu[1]

[1] *Department of Electrical and Electronics Engineering, AYBU, Ankara, Türkiye*

***Contact:** yildirim.makif20@gmail.com, phone: +90-312-906-2217

***Abstract***—The rapid advancement of quantum computing poses a significant threat to classical public-key cryptographic systems, necessitating the transition to Post-Quantum Cryptography (PQC). This study investigates the implementation challenges of NIST-standardized signature schemes on resource-constrained embedded hardware. We present a comparative analysis of SPHINCS+ and CRYSTALS-Dilithium on an ARM Cortex-M4 (STM32F407G) microcontroller. Our findings reveal that SPHINCS+ is practically unusable in this software-only environment, with impractical execution times. Furthermore, the reference Dilithium implementation failed to execute entirely on the MCU due to severe RAM and timing constraints. To overcome these hardware limitations, we integrated a hardware-accelerated Dilithium core onto a Xilinx Zynq-7000 ZedBoard SoC. By implementing a specialized Number Theoretic Transform (NTT) accelerator in the FPGA fabric, we achieved successful execution with performance rates for key generation and signature generation at millisecond levels. These results demonstrate that while pure software PQC is non-viable for standard microcontrollers, a hardware-software co-design approach provides the necessary efficiency for quantum-resistant embedded systems.

## I. Introduction

Large-scale Quantum Computing has the potential to pose a threat to the current global cryptographic infrastructure, as theorized by Nielsen and Chuang [1]. The mathematical backbone of RSA and ECC can be solved via Shor's algorithm utilizing integer factorization and discrete logarithm problems in polynomial time by a powerful quantum processor [2], [3], [8]. This vulnerability has caused a paradigm shift toward Post-Quantum Cryptography (PQC), led by the National Institute of Standards and Technology (NIST) in 2016 [4].

In this paper, we focused on two digital signature schemes, CRYSTALS-Dilithium and SPHINCS+. These methods have different mathematical approaches: lattice-based and hash-based cryptography [5], [6]. These algorithms are resilient against "Store Now, Decrypt Later" (SNDL) attacks for post-Quantum era.

Nowadays, global cryptographic infrastructure relies on embedded solutions. PQC methods, on the other hand, are a new challenge for embedded systems. Platforms such as the ARM Cortex-M architecture often lack the necessary hardware requirements to manage high-degree polynomial multiplications of Dilithium or hypertree structures of SPHINCS+ [7].

This paper presents a comparative performance analysis of these NIST standards. We first documented the problems encountered when attempting software-only implementations on an STM32F407G MCU, where SPHINCS+ latencies reached around 10 minutes and Dilithium completely failed due to memory restrictions. Then, we demonstrated a successful new design based on hardware-software co-design, utilizing the Xilinx Zynq-7000 SoC for hardware acceleration [12]. Our results prove that for real-time embedded applications, hardware acceleration is not a mere optimization but a functional necessity.

This paper is organized as follows: Section II reviews the background of lattice-based and hash-based signature schemes, along with related work about hardware acceleration. Section III details the methodology and experimental setup for both the MCU and SoC platforms. Section IV provides the implementation details, and Section V presents the comparative results and discussion. At the end, Section VI concludes the paper.

## II. Foundations of PQC and Hardware Implementations

This section outlines the mathematical foundations of the selected NIST standards, CRYSTALS-Dilithium and SPHINCS+, focusing specifically on the computational bottlenecks they introduce. Furthermore, it reviews the limitations of resource-constrained environments and the current landscape of hardware acceleration, highlighting why System-on-Chip (SoC) architectures are suited to address these cryptographic constraints.

### A. CRYSTALS-Dilithium and Lattice-Based Security

CRYSTALS-Dilithium is a digital signature scheme based on the Module Learning with Errors (MLWE) problem [5]. Its security is derived from the shortest vector problem in lattices, which remains hard for both classical and quantum computing capabilities. The primary bottleneck in Dilithium is the multiplication of high-degree polynomials, typically optimized using the Number Theoretic Transform (NTT) to reduce complexity [12]. Despite these optimizations, the memory footprint remains significant; as documented in the NIST specifications, even Level 2 security requires substantial stack space for coefficient storage [5]. Fig. 1 is a schematic presenting the application steps of the CRYSTALS-Dilithium scheme.

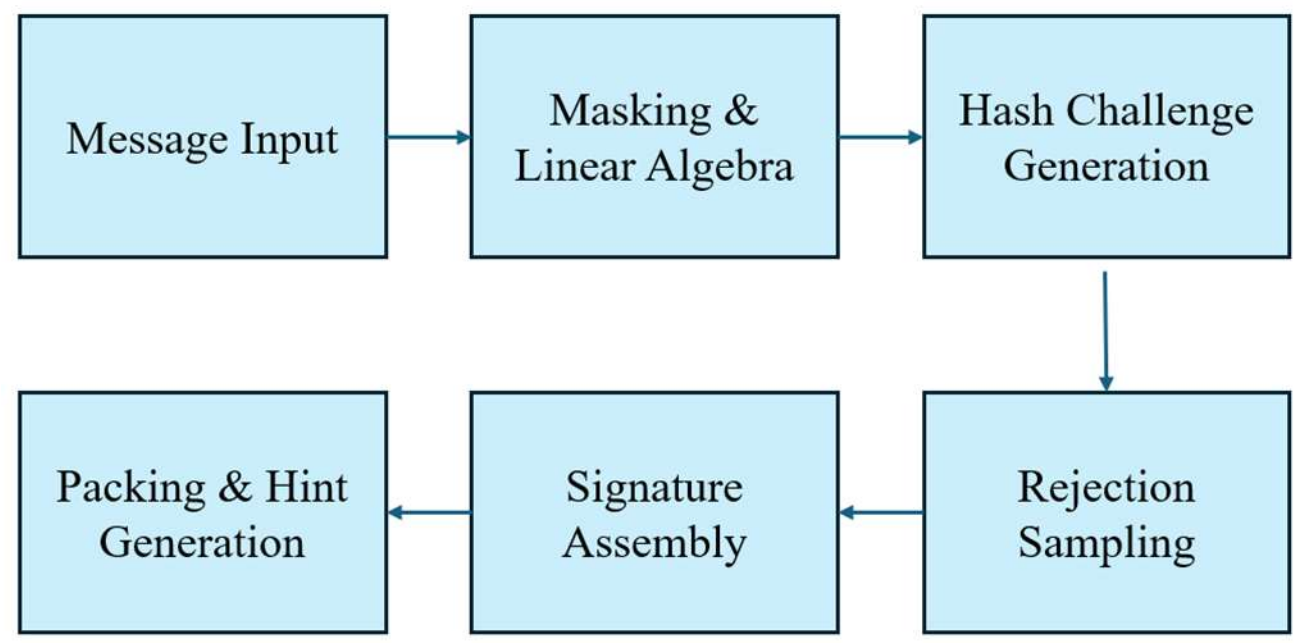

**Fig. 1** Dilithium signing procedure.

The CRYSTALS-Dilithium signing process starts by taking the message and the private key, applying a random mask, and using the Number Theoretic Transform (NTT) to perform matrix-vector multiplications that hide the key. This masked data is combined with the message and hashed to generate a challenge string. To prevent the secret key from leaking, the algorithm uses rejection sampling—if the coefficient sizes fall outside strict security bounds, it throws them out and restarts the masking step. Once an iteration passes this check, the valid vectors are assembled into the signature and bit-packed with cryptographic "hints" so the verifier can reconstruct the matrices without needing the full uncompressed data.

### B. *SPHINCS+ and Stateless Hash-Based Frameworks*

SPHINCS+ is a stateless hash-based signature scheme, evolving from the original SPHINCS design to improve security, reduce signature sizes, and optimize parameter sets for various performance trade-offs [6], [10]. It relies on a Merkle tree structure composed of Winternitz One-Time Signatures (WOTS+) and Forest of Random Subsets (FORS) few-time signatures [9]. Its primary advantage is its minimal security assumptions—requiring only a secure hash function—but this comes at the cost of high computational density. Research by Niederhagen et al. has explored "streaming" signatures to optimize memory constraints, yet the signing process remains prohibitively slow on low-power microcontrollers [7].

The SPHINCS+ signing process starts by heavily hashing the input message to create a digest and an index, which determines the exact path through its hyper-tree structure. The message digest is first signed at the bottom layer using the Forest of Random Subsets (FORS) scheme. The public key outputs from FORS are then signed using the Winternitz One-Time Signature (WOTS+) algorithm. Finally, the algorithm calculates the Merkle tree authentication paths to link these WOTS+ signatures all the way up to the main root node. The final signature package simply contains the FORS signature, the WOTS+ signatures, and the full sequence of authentication paths needed to verify it against the globally trusted public key.

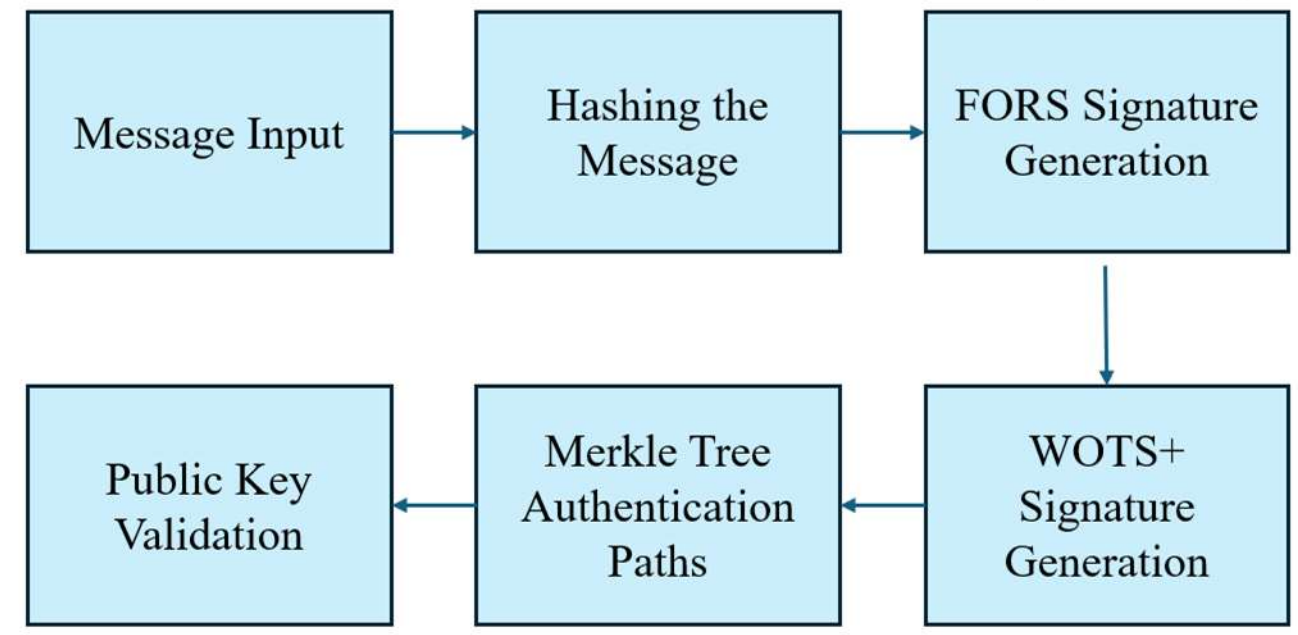

**Fig. 2** SPHINCS+ signing procedure.

### C. *Challenges in Resource-Constrained Environments*

The use of PQC on MCUs is a rapidly growing field of research. Previous studies on ARM Cortex-M4 platforms have shown that while verification can often be performed within acceptable time bounds, key generation and signing frequently exceeded the capabilities of general-purpose MCUs [7]. Memory-restricted devices face a particular challenge with Dilithium, storing both the operating system/application stack and the large intermediate buffers required for polynomial arithmetic in their SRAMs.

### D. *Hardware Acceleration and the ZYNQ Architecture*

Today’s requirements demand not only high performance but also strong security guarantees, making the use of hardware accelerators increasingly essential in modern systems. Computationally intensive cryptographic operations such as hashing and Number Theoretic Transform (NTT) benefit significantly from parallel execution capabilities that are difficult to achieve efficiently on general-purpose processors alone. In this context, the use of Field Programmable Gate Arrays (FPGAs) provides a highly flexible and powerful solution for mitigating the bottlenecks of both lattice and hash-based schemes [13], as they enable the design of specialized hardware architectures tailored for parallelization and pipelining of these critical operations.

FPGAs allow designers to exploit fine-grained parallelism and optimize data paths, resulting in substantial improvements in throughput and latency. A notable example is the recent work on the CityUHK-CALAS architecture, which demonstrates an effective partitioning strategy between the Programmable Logic (PL) and the Processing System (PS) within a System-on-Chip (SoC) platform such as the Xilinx Zynq-7000. In this approach, low-level computational primitives, including hashing and NTT operations, are offloaded to the PL, where they can be executed in parallel with high efficiency. Meanwhile, higher-level protocol logic and control mechanisms are handled by the PS, which provides greater flexibility and ease of implementation.

This division of responsibilities not only leverages the strengths of both hardware and software components but also significantly reduces overall system latency. By carefully balancing the workload between PL and PS, the architecture achieves a more efficient utilization of resources and enhances performance without compromising security. Consequently, such co-design approaches represent a promising direction for developing optimized platforms capable of meeting the stringent requirements of next-generation secure and high-performance applications [12].

## III. Methodology and Experimental Setup

The assessment of NIST PQC standards was conducted across two distinct branches of hardware. This dual-phase approach revealed the performance complications on general purpose MCUs and the major gains on the specialized hardware acceleration.

### *A. Microcontroller Application: STM32F407G*

In this stage, the STM32F407G-DISC1 board was utilized to evaluate an affordable embedded solution. The processor of this platform is 32-bit ARM Cortex-M4 core operating at 168 MHz. The memories of this microcontroller are 192 KB SRAM and 1 MB of FLASH memory.

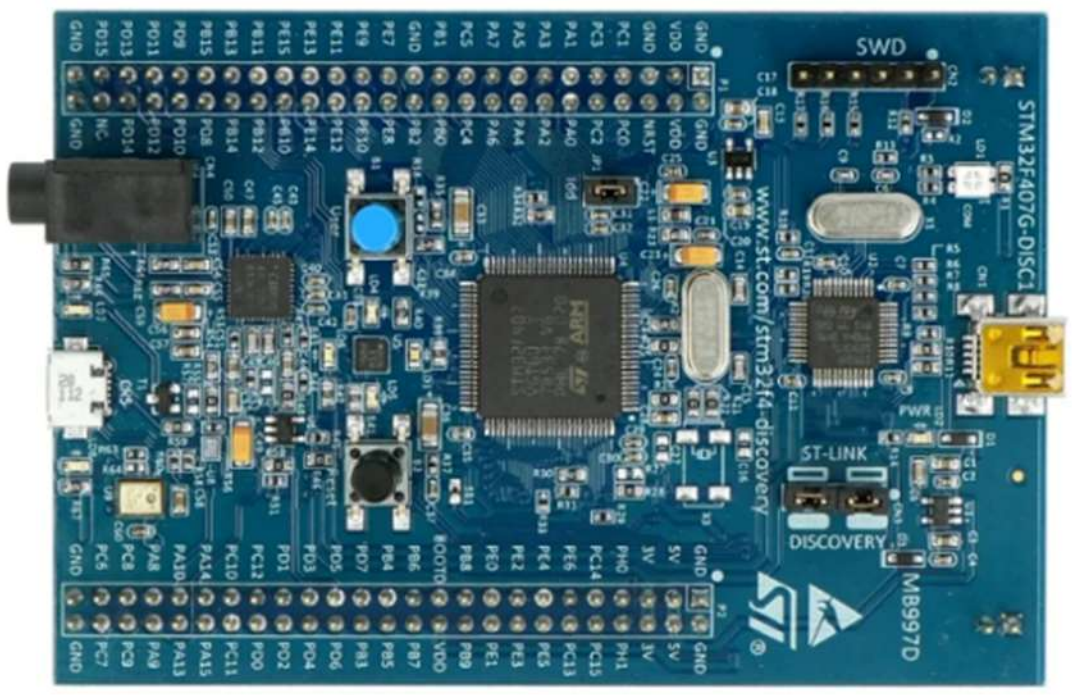

**Fig. 3** STM32F407G-DISC1 Board.

Target algorithms implemented here are the reference implementations for CRYSTALS-Dilithium and SPHINCS+ to establish a standard-compliant baseline. The objective was to assess the feasibility of the unmodified, software-only algorithms within strict memory boundaries, without applying any hardware-specific optimizations or algorithmic simplifications. The board that has been used in the setup can be seen in Fig. 3. This setup allows for a realistic evaluation of how such algorithms would perform in practical embedded environments under constrained resource conditions.

### *B. HW-SW Co-Design: ZYNQ-7000 SoC*

In order to overcome the limitations encountered in the previous phase, the project was migrated to the Xilinx Zynq-7000 ZedBoard (XC7Z020) platform, which is illustrated in Fig. 4.

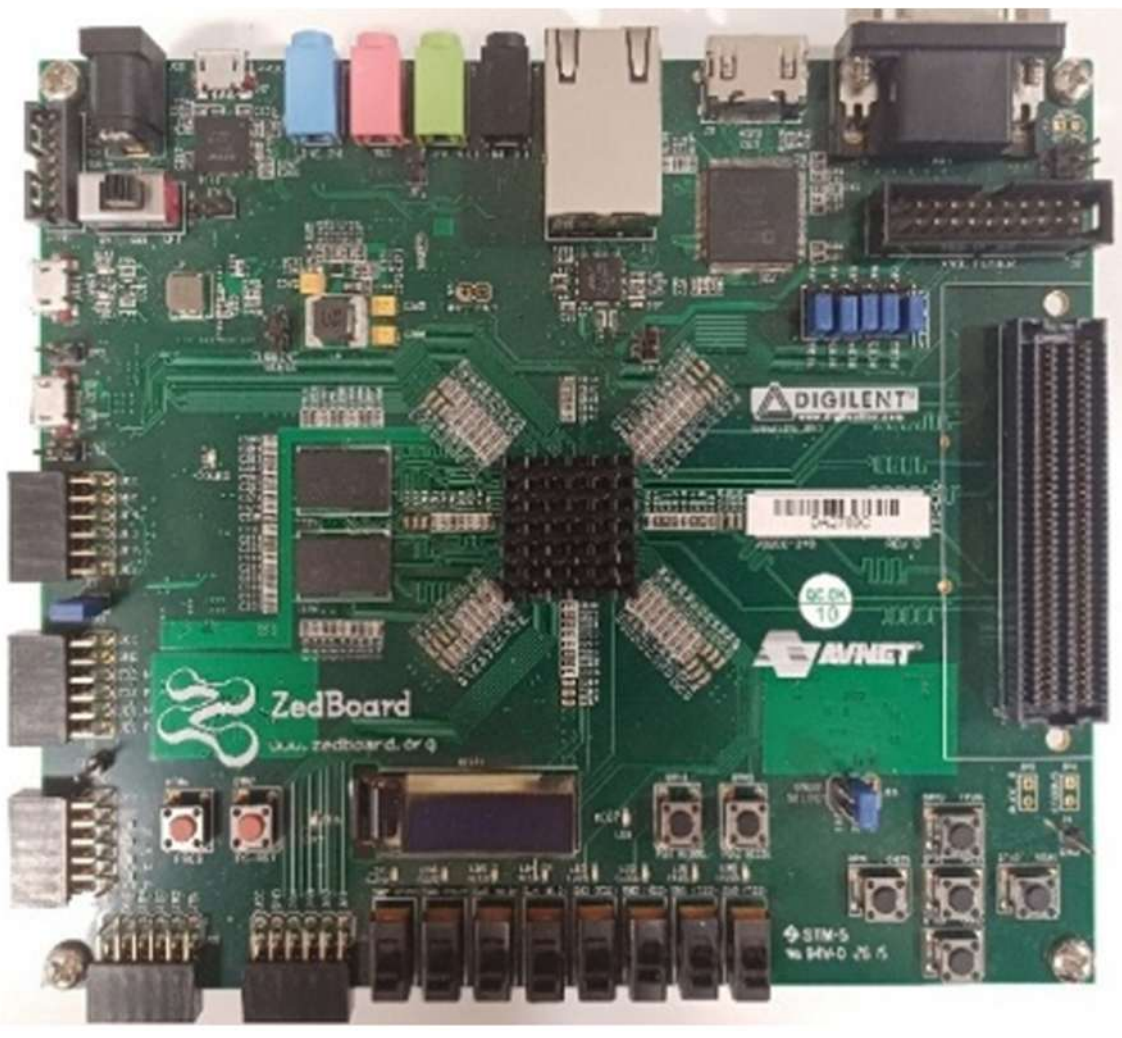

**Fig. 4** Zedboard ZYNQ 7000 Board.

The architecture of this platform provides a Processor System (PS) –a dual core ARM Cortex-A9– and a Programmable Logic (PL) equivalent to an Artix-7 FPGA fabric. We adopted an architecture called CityUHK-CALAS, a high performance configurable HW-SW co-design developed for CRYSTALS-Dilithium [12]. The design has been partitioned as the PL to target the offloading of the computationally heavy side of the CRYSTALS-Dilithium like Number Theoretic Transform (NTT), Inverse NTT (INTT) and Keccak-based hashing operations. The PS was aimed at managing memory constraints and higher-level requirements.

## IV. Implementation Details

This section details the execution of both testing phases. First, we examine the specific memory and timing bottlenecks encountered when porting the unmodified PQC algorithms to the bare-metal STM32 microcontroller. Following this failure analysis, we detail the hardware-software co-design strategy on the Zynq-7000, explaining how the AXI protocols were configured to successfully offload the heavy arithmetic from the ARM processor to the FPGA fabric.

### *A. Analysis of Software Implementation Failures on STM32*

Porting the NIST standards to the Cortex-M4 revealed some mismatches between hardware resources and heavy algebraic requirements of PQC methods. Due to this it is shown that most of the standard microcontrollers cannot be used. It is also noteworthy that STM32 has a list of cryptography methods that are implementable for STM platforms [11].

*Stack Overflow (CRYSTALS-Dilithium):* Dilithium relies on the Module Learning with Errors (MLWE) problem, which requires large local arrays for polynomial coefficients and intermediate transform buffers during key and signature generation [5]. On the other hand, STM32F4 had 192KB of SRAM which was not enough for Dilithium. It is noteworthy that algorithms working with the on-the-fly principle to reduce the RAM requirements are excluded.

*Computational Latency (SPHINCS+):* SPHINCS+ utilizes a Merkle tree structure composed of Winternitz One-Time Signatures (WOTS+) and Forest of Random Subset (FORS) [6]. The scheme successfully executed through signature and key generation without exceeding memory limits taking around 10 minutes for each step. On the verification process requiring thousands of hash function evaluations, the system collapsed. Lacking the dedicated hardware for hashing functions caused the SPHINCS+ execution to fail during the verification step. These findings prove that STM32F4 cannot run PQC algorithms as an embedded system, which is also shown on the website of ST Microelectronics [11].

### *B. FPGA Integration and AXI Logic*

The successful acceleration on the ZYNQ-7000 relies on a module parallelization of the arithmetic primitives using the CityUHK-CALAS core [12].

Modular Acceleration: The hardware core features a highly optimized butterfly unit for NTT/INTT polynomial multiplication alongside a dedicated PRNG module with an SHA-3/Keccak core for sampling coefficients.

Inter-Processor Communication: To prevent data bottlenecks and synchronization issues between PS and PL, AXI protocols were used. The AXI4-Lite interface was used for transmitting control signals, while high-performance AXI4-Stream ports were used for the data transfer of public keys, messages and signature buffers.

Execution Flow: The system acts as a true co-processor model. The ARM core executes the top-level state machine, handling message formatting and high-level protocol logic. The FPGA side is triggered only when intensive mathematical blocks are required. Thereby, the latency of complex lattice arithmetic is masked.

Fig. 5 illustrates the top-level hardware-software co-design architecture implemented on the Zynq-7000 SoC [12]. The system is distinctly partitioned into two domains: the Processing System (PS) and the Programmable Logic (PL). The ARM-based PS acts as the master controller, managing the top-level state machine and message formatting. The computationally intensive cryptographic primitives—specifically the NTT/INTT polynomial multipliers and the Keccak/SHAKE hashing modules—are offloaded to the PL fabric as a dedicated hardware accelerator. To ensure efficient communication without bottlenecking the system, the architecture utilizes Advanced eXtensible Interface (AXI) protocols. AXI-Lite is employed for lightweight control signals and register configuration, while high-performance AXI-Stream interfaces, managed via Direct Memory Access (DMA), handle the rapid transfer of bulk data such as public keys and signature buffers.

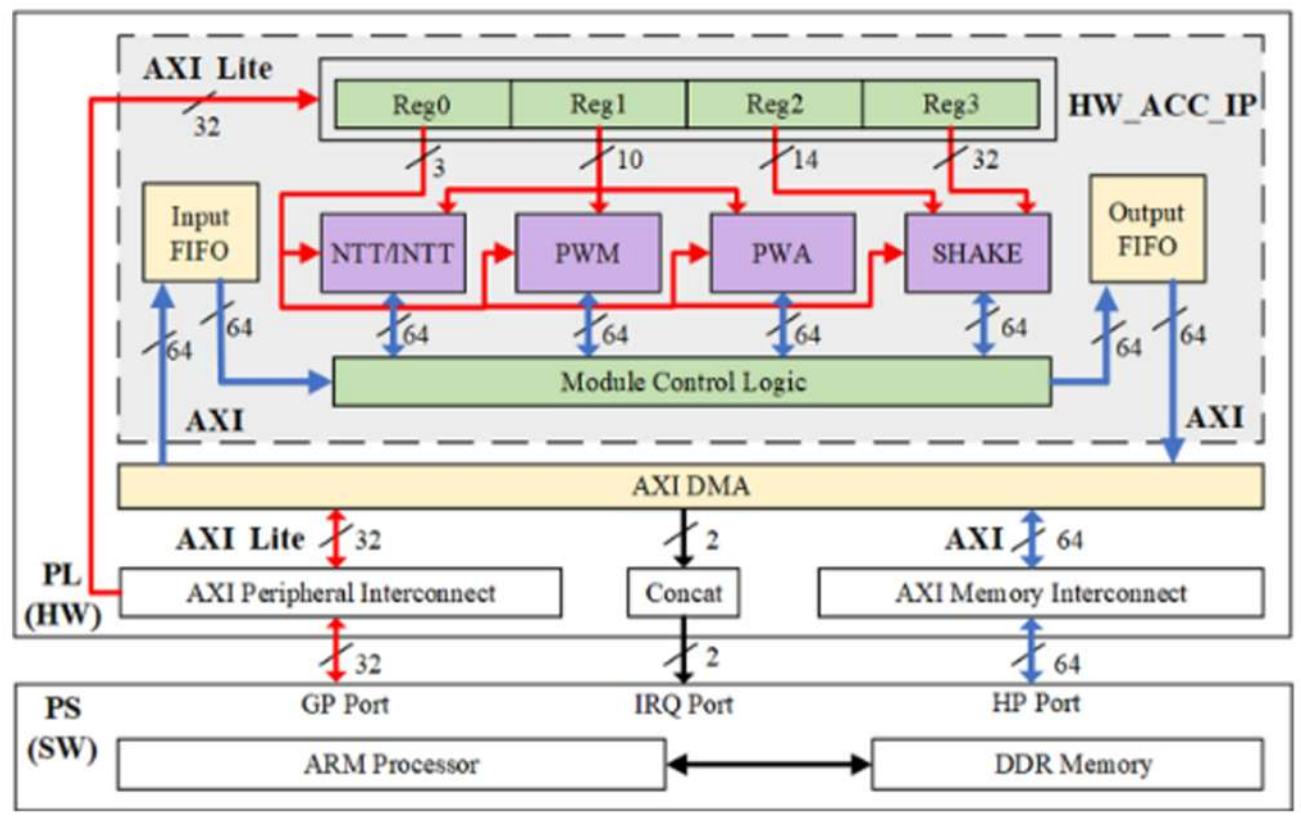

**Fig. 5** The top-level SW/HW co-design SoC architecture.

## V. RESULTS AND PERFORMANCE ANALYSIS

The evaluation and results clearly show that the operational boundaries of microcontroller-based PQC favor performance advantages of hardware acceleration. The metrics presented reflect the average execution times observed during testing.

As shown in the detailed results in Table I, both algorithms encountered critical bottlenecks related to hardware architecture.

**TABLE I**
PERFORMANCE RESULTS FOR STM32F407G-DISC1

| PQC Method | Duration of | | |
|---|---|---|---|
| | Key Generation | Signature Generation | Verification |
| SPHINCS+ | ~10 minutes | ~10 minutes | N/A |
| Dilithium | N/A | N/A | N/A |

The failure of CRYSTALS-Dilithium is directly caused by its MLWE foundation and 192 KB SRAM capacity. The memory footprint rapidly exceeded the SRAM capacity of the STM32F4 and resulted in a stack overflow situation at the initial stages. Also, while SPHINCS+ is a stateless scheme that avoids massive polynomial arrays, its reliance on thousands of iterative hash function calls for Merkle tree created a massive computational bottleneck [6]. Lacking hardware-level hashing acceleration, the Cortex M4 required approximately 20 minutes to generate signatures and keys, then collapsed during verification.

The migration to the ZYNQ-7000 SoC eliminates the memory overflows and computational bottlenecks observed in Phase 1. With an optimized approach, excellent results have been achieved compared to STM32F4. NTT and Keccak hashing have been offloaded to the FPGA fabric via the CityUHK-CALAS architecture [12], resulting in the results given in Table II.

**TABLE II**
PERFORMANCE RESULTS FOR ZEDBOARD – ZYNQ-7000

| PQC Method | Duration of | | |
|---|---|---|---|
| | **Key Generation** | **Signature Generation** | **Verification** |
| Dilithium | 1.109 ms | 5.94 ms | 1.17 ms |

The results of this study clearly demonstrate that standard microcontroller architectures are under-equipped to handle the mathematical density and complexity of PQC methods. The STM32F407G suffered from immediate stack overflows during Dilithium execution due to its 192 KB SRAM limit, indicating that the available memory is insufficient for handling the intermediate data structures required by lattice-based computations. Similarly, the hash-heavy SPHINCS+ algorithm resulted in impractically long execution times of approximately 10 minutes, making it unsuitable for time-critical embedded applications.

On the other hand, offloading the Number Theoretic Transform (NTT) and Keccak hashing operations to the PL on the Zynq-7000 SoC radically changed system performance. By mitigating the computational bottlenecks of lattice-based schemes, execution times were reduced to the millisecond range. This stark performance gap indicates that software-only PQC implementations are insufficient for resource-constrained devices. Emerging quantum threats necessitate the adoption of these complex algorithms, and hardware dedicated cryptographic processors will become a necessity for secure embedded and larger systems.

## VI. CONCLUSION

This study investigated the practical challenges associated with migrating to NIST Post-Quantum Cryptography standards within resource-constrained embedded systems. Our baseline implementation on the STM32F407G microcontroller demonstrated that conventional 32-bit architectures are vulnerable to issues such as stack overflows in lattice-based schemes like CRYSTALS-Dilithium and high computational latency in hash-based schemes such as SPHINCS+. These findings indicate that purely software-based approaches on low-power embedded platforms are insufficient to meet the stringent performance and reliability requirements of next-generation cryptographic standards. To overcome these limitations, a second implementation was performed on the Xilinx Zynq-7000 System-on-Chip, enabling a hardware-software co-design approach in which computationally intensive primitives—particularly the Number Theoretic Transform (NTT) and Keccak hashing functions—were offloaded to the FPGA fabric for parallel and pipelined execution, while higher-level protocol logic was handled by the processing system. This architecture effectively leveraged the strengths of both domains, eliminating memory bottlenecks and significantly reducing execution latency to the millisecond level. Overall, the results highlight the necessity of heterogeneous computing architectures for efficient deployment of post-quantum cryptographic algorithms in embedded systems, demonstrating that FPGA-accelerated solutions provide a scalable and high-performance alternative to traditional microcontroller-based implementations. As a direction for future work, further optimization of FPGA resource utilization and power consumption, extension of the framework to additional post-quantum schemes, and evaluation of side-channel attack resistance are important steps toward achieving more robust and practical implementations.